# Energy Efficient Stochastic Signal Manipulation in Superparamagnetic Tunnel Junctions via Voltage-Controlled Exchange Coupling

*Qi Jia, Onri J. Benally, Brandon Zink, Delin Zhang, Yang Lv, Shuang Liang, Deyuan Lyu, Yu-Chia Chen, Yifei Yang, Yu Han Huang, and Jian-Ping Wang\**

ABSTRACT: Superparamagnetic tunnel junctions (sMTJs) are emerging as promising components for stochastic units in neuromorphic computing, owing to their tunable random switching behavior. Conventional MTJ control methods, such as spin-transfer torque (STT) and spin-orbit torque (SOT), often require substantial power. Here, we introduce the voltage-controlled exchange coupling (VCEC) mechanism, enabling switching between antiparallel and parallel states in sMTJs with an ultralow power consumption of only 40 nW, approximately two orders of magnitude lower than conventional STT-based sMTJs. This mechanism yields a sigmoid-shaped output response, making it ideally suited for neuromorphic computing applications. Furthermore, we validate the feasibility of integrating VCEC with the SOT current control, offering an additional dimension for magnetic state manipulation. This work marks the first practical demonstration of VCEC effect in sMTJs, highlighting its potential as a low-power control solution for probabilistic bits in advanced computing systems.

TEXT:

Neuromorphic computing circuits built on CMOS transistors are constrained by large circuit areas and high energy demands, as the deterministic nature of transistors is inherently misaligned with the requirements of neuro-mimetic algorithms[1–3]. By directly mimicking the neural and synaptic unit from at the hardware level using manipulatable stochastic units, computational costs can be significantly reduced through approaches like stochastic computing[1,4–8]. Superparamagnetic tunnel junctions (sMTJs) emerge as a promising candidate for elementary stochastic unit cells, offering potential reductions in both area and energy consumption[1,3,9–19]. While recent efforts on sMTJ have primarily focused on maximizing the speed of individual sMTJ to generate stochastic bits at GHz or higher frequencies[20–26], little attention has been given to enhancing the energy efficiency of sMTJ manipulation. The situation is related to the bottleneck of increasing the efficiency of the control mechanisms, including external magnetic fields, spin-transfer torque (STT)[27–31], or spin-orbit torque (SOT)[2,32–35]. All these methods require current densities exceeding $\sim 10^6$ A/cm$^2$ to generate a sufficient Oersted field or spin torque, resulting in considerable energy consumption. Due to inevitable fabrication inconsistencies among sMTJs, achieving consistent output across large sMTJs arrays in an energy-efficient way is critical for practical applications. Therefore, an energy-efficient method to control the stochastic behavior of sMTJs, such as utilizing voltage-induced effects, is highly desirable.

Voltage-controlled magnetic anisotropy (VCMA), a well-known, low power, voltage-induced effect has been explored for modulating the stochastic behavior of sMTJs in artificial neural network (ANN) applications[36–41]. However, VCMA only modulates the switching rate by altering the energy barrier, without impacting the MTJ output level. To overcome this limitation, it is essential to have a voltage effect that can generate a bipolar magnetic field. In our previous work, we presented evidence showing that the strength of the Ruderman-Kittel-Kasuya-Yosida (RKKY) coupling in ferromagnetic/non-magnetic/ferromagnetic/insulator (FM/NM/FM/I) structures can be modulated by the applied voltage, a phenomenon termed as voltage-controlled exchange coupling (VCEC)[42–44]. The effect on the free layer of the MTJ is analogous to STT, where the electric field of different directions favors either an antiparallel (AP) or parallel (P) configuration. Remarkably, the reported critical current density for this effect, measured by conductive atomic force microscopy (C-AFM), is as low as $1.1\times 10^5 A/cm^2$, an order of magnitude lower than that of the most efficient STT-MTJ[42]. While this concept shows promise, VCEC effect alone has yet to be demonstrated for bipolar MTJ manipulation at the device level.

In this work, we fabricate perpendicular sMTJ with the free layer antiferromagnetically coupled to the underneath layer via interlayer exchange coupling. The presence of the VCEC effect in our structure is confirmed by observing an opposite switching direction compared to STT. Time-domain measurement reveal that the output level can be modulated by both the external magnetic field and applied voltage, following a sigmoid function. The use of the VCEC mechanism effectively reduces the energy consumption to 40 nW. The switching rate of sMTJ is observed to be influenced by the accompanying VCMA effect, with negligible thermal impact. Additionally, integrating the current through the underlying SOT channel introduces an additional control dimension, evolving the original dual-biasing scheme into a tri-biasing

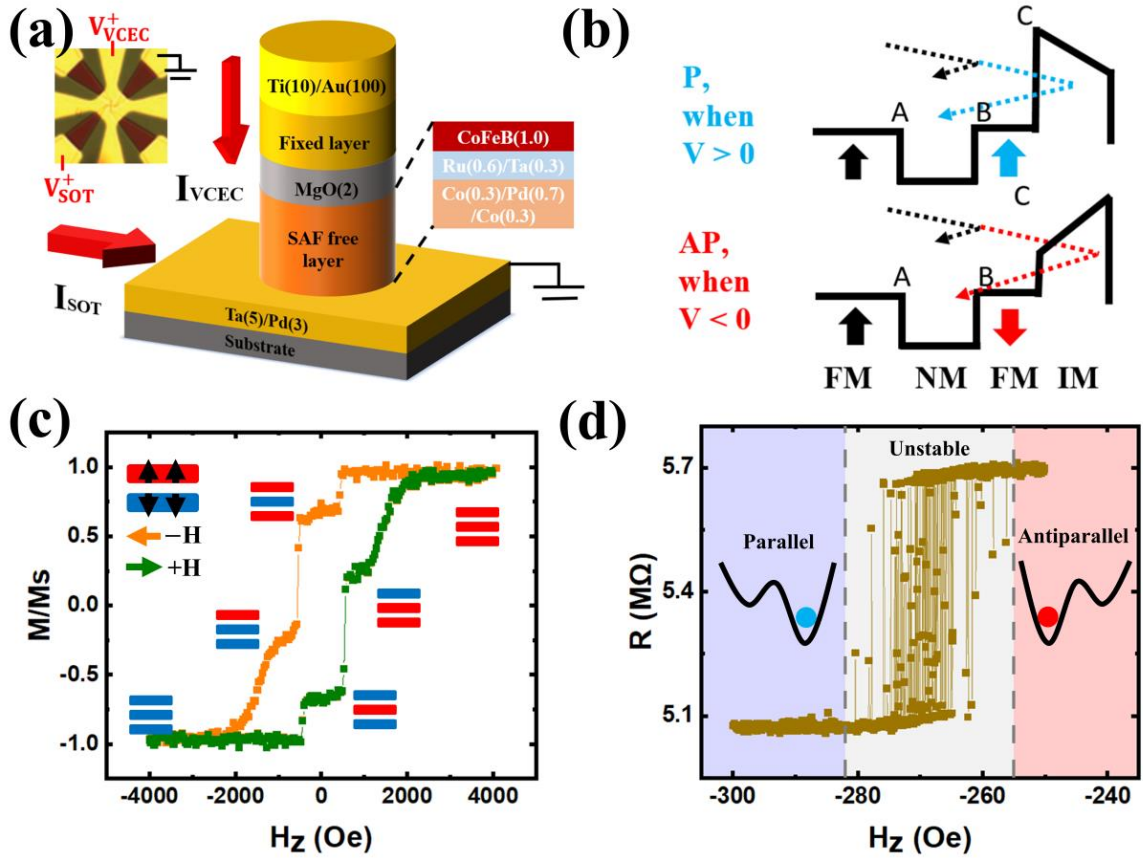

Figure. 1. Stack structure and device properties. (a) Schematic of the MTJ device structure and current flow directions, with a top-view microscope image inset (top left). (b) Diagram of the voltage-controlled exchange coupling (VCEC) mechanism, showing spin reflection at interface C, modulated by an applied voltage to control the free layer's magnetization direction. (c) Post-annealing M-H loops of the stack, displaying sharp switching behavior, indicative of perpendicular magnetic anisotropy (PMA) in all three magnetic layers. The red and blue blocks represent the out-of-plane magnetization of each layer, with the red block pointing upwards and the blue block pointing downwards. (d) R-H minor loop of a 100 nm MTJ diameter pillar, demonstrating PMA retention as the free layer oscillates between anti-parallel (AP) and parallel (P) states in the unstable region.

scheme.

The structure of the MTJ is shown in Fig. 1(a). The bottom free layer CoFeB is coupled antiferromagnetically with [Co (0.3 nm)/Pd (0.7 nm)]$_3$ through Ru (0.6 nm)/Ta (0.3 nm) via the RKKY interaction. The top CoFeB serves as the reference layer and is pinned by [Pd (0.7 nm)/Co (0.3 nm)]$_8$ through the Ta spacer. To ensure the dominance of the voltage effect rather than those induced by current, a thick MgO layer of 2 nm is grown. Applying a voltage across the MTJ pillar is expected to modify the interlayer exchange coupling through the VCEC effect, as illustrated in Fig. 1(b). The applied voltage modulated the spin-dependent reflectivity of the electron wave function at the FM/MgO interface, which altered the ground state between AP and P states. Fig. 1(c) shows the out-of-plane M-H loop of the stack after annealing. The three-step switching (for magnetic field from one end to the other) corresponds to the switching of the CoFeB free layer, the bottom [Co (0.3 nm)/Pd (0.7 nm)]$_3$ coupled with the CoFeB bottom free layer and the top CoFeB layer, respectively. In the electrical study, the range of the magnetic field is controlled so that only the CoFeB free layer is switched during the electrical measurement. Fig.1(d) shows the R-H minor loop measured under a small current of 5 nA, with a magnetic field sweep rate of approximately 2 Oe/s. The stochastic resistance switching between -255 Oe and -280 Oe indicates that the MTJ being measured is in a superparamagnetic state while maintaining the perpendicular anisotropy. The large resistance-area product RA of $4.01 \times 10^4 \; \Omega \cdot \mu m^2$ is consistent with the thick MgO barrier. The tunnel magnetoresistance is determined to be 12.9 % (TMR = $\frac{R_{AP} - R_P}{R_P}$).

To study the stochastic behavior in detail, the circuit shown in Fig. 2(a) was developed to analyze the stochastic signal of the sMTJ. A constant current of different values is applied through the MTJ for the bias-dependent measurement. The time-dependent voltage is measured after removing the DC component with a bias tee and amplified. The strategy of applying a current is designed to allow for the separation of any effects induced by the SOT current, as the current source can maintain a constant voltage across the pillar even when adding current through the bottom layer. Given that the TMR value is around 12.9%, it is reasonable to assume the voltage does not significantly change before and after the switching. The applied current can be converted to the average voltage of the two states interchangeably. The detailed V - I curve is shown in Fig. S1. Example waveforms under different conditions are shown in Fig. 2(b). From the acquired waveforms, the dwell time, defined as the time interval between successive switching, can be extracted. The histogram of the dwell time between AP and P states over a 40-second waveform under condition of I = 200 nA and H = -259 Oe is shown in Fig. 2(c). The dwell time follows an exponential distribution, consistent with the time distribution expected from a Poisson process.

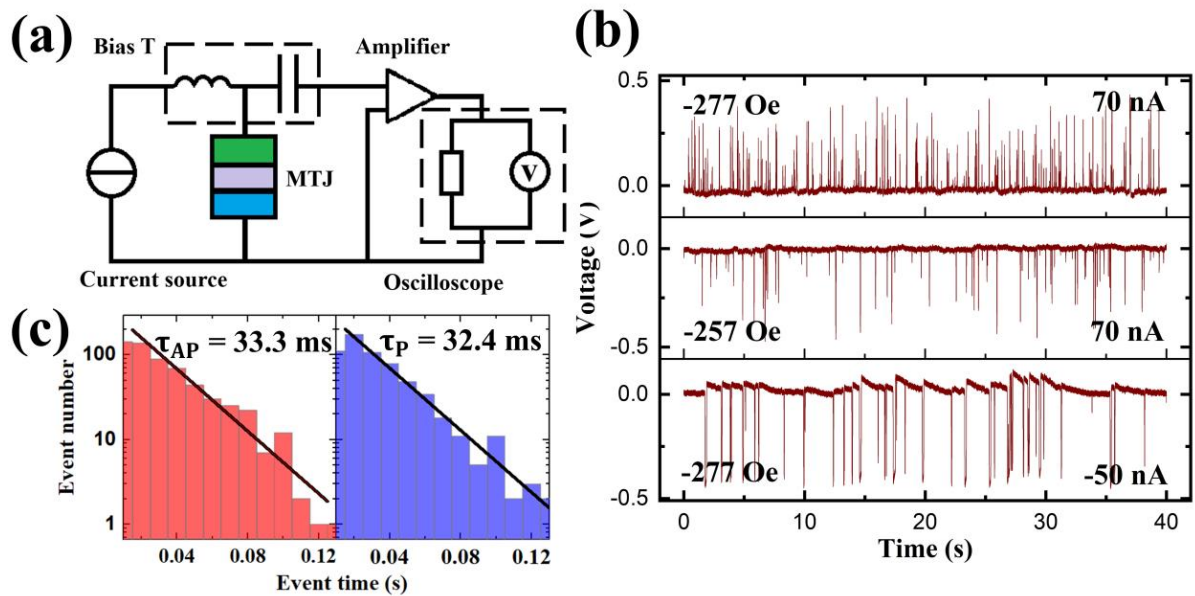

Figure 2. Time domain measurement of the random telegraph noise (RTN) of sMTJ. (a) Circuit schematic for time-resolved measurement of the MTJ state. (b) Histogram of the dwell time in the AP and P states under condition of I = 200 nA and H = -259 Oe. (c) Example measured waveforms of 40 s showing output level changes in response to variations in the external magnetic field or applied current.

In a two-state system composed of two Poisson processes, the transition probability over a given time period depends on the height of the energy barrier. In an MTJ system with coherent rotation of the free layer, the dwell time of each state under the external magnetic field can be quantitatively described by the Néel-Brown formula:

$$\tau_{AP,P} = \tau_0 \exp\left[\frac{E_b}{k_B T}\left(1 \pm \frac{H_Z - H_s}{H_K^{eff}}\right)^{n_H}\right], \quad (1)$$

where the $\tau_0$ is the attempt frequency (usually around 1ns), $E_b$ is the barrier height, $H_K^{eff}$ is the effective magnetic anisotropic field, $n_H$ is assumed to be 2 by assuming a pitchfork bifurcation energy landscape[45], $H_s$ is the effective stray field experienced by the free layer and $H_Z$ is the external magnetic field. The interlayer exchange coupling magnetic field is included in $H_s$ to simplify the expression ($H_s(V) = H_{s,real} + H_{ex}(V)$). The sign of $H_Z$ is positive when H favors the AP state.

If we define AP rate as the possibility of finding the MTJ in the AP state $P_{AP}$, then

$$P_{AP} = \frac{\tau_{AP}}{\tau_{AP} + \tau_P}. \quad (2)$$

By substituting (1) into (2), we get

$$P_{AP} = \frac{1}{1 + \exp\left(-4\frac{E_b}{k_B T}\frac{H_Z - H_s}{H_K^{eff}}\right)}. \quad (3)$$

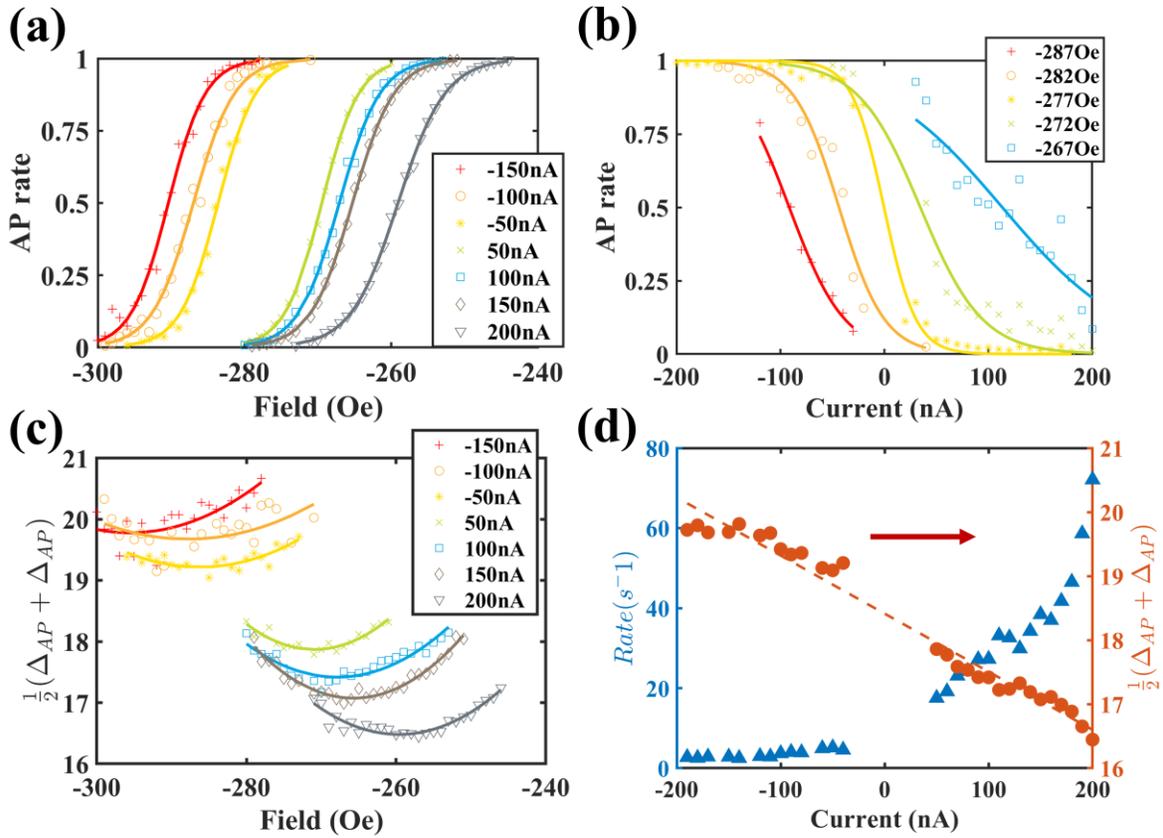

Figure 3. Magnetic field and voltage Influence on AP rate and switching rate. The anti-parallel (AP) switching rate of the stochastic signal manipulated by (a) the external magnetic field and (b) VCEC, correlated with the applied current. (c) Variation in average stability as a function of the magnetic field under different current levels. (d) Minimum average stability and corresponding switching rate as a function of the applied current.

The experimental result of AP rate as a function of the perpendicular external magnetic field under different currents is plotted in Fig. 3(a). For each current, the MTJ transitions to the AP state under a positive magnetic field display a sigmoid function, consistent with Eq. (3). This behavior suggests that the sMTJ could be utilized to directly mimic neural excitation at the hardware level. The expression also enables measurement of changes in the exchange coupling magnetic field ($\Delta H_{ex}$) from the VCEC effect. In Fig. 3(a), The shift of the sigmoid excitation curve along the magnetic field axis under different currents indicates an additional equivalent magnetic field induced by VCEC. Notably, such a bipolar shift cannot be achieved by VCMA alone. The strength of VCEC is thus estimated to be -175 Oe/μA (approximately -25 Oe/V). Compared with traditional measurements using a single R-H minor loop, our AP rate measurement achieves lower error value due to the increased number of switching events within a comparable measurement duration. The magnetic field at AP rate = 50% switching point is extracted as the reference point through fitting, ensuring that variations in the coefficient $\frac{E_b}{k_B T \cdot H_K^{eff}}$ under different currents do not introduce additional error.

In Fig. 3(b), the effect of the applied current with fixed magnetic fields is shown. The range of the current is fixed within ± 200 nA to stay in the direct tunneling region and avoid the breakdown (see Fig. S1(d)). The noisy points around the zero-current come from the larger noise signal ratio and the corresponding larger error of the measured dwell time. We observed that the AP rate is manipulated unidirectionally like the STT effect. The sigmoid function shape validates that the VCEC-induced effective magnetic field could be used to excite the MTJ stochastic unit. The behavior can be also described by Eq. (3) while the change of the exchange coupling is affecting the $H_s$ by $H_{ex}$ with a fixed $H_Z$ value. Interestingly, when the magnetic field is fixed to -277 Oe, the MTJ state could be switched from AP to P within ± 100 nA, corresponding to a current density of ±1.27×10$^4$ A/cm$^2$, smaller than previously reported VCEC switching current density in a similar structure. The power corresponding to the maximum switching current is as low as 40 nW. As a comparation, the power for STT-based MTJ is around 10 μW (supplementary note 1). In the previous report, the effect was attributed to VCEC instead of STT by assuming that the small current density is insufficient to generate any observable STT effect[42]. In our simplified MTJ with a single SAF structure, a solid argument ruling out the contribution of the STT effect comes from the analysis of switching direction. Since the pinning layer is at the top, a positive current (electrons tunneling from the free layer to the pinned layer) should favor the anti-parallel (AP) state in the STT scenario. The observed effect, with inverted polarity, successfully rules out the STT contribution and robustly demonstrates the existence of the VCEC effect. Additional evidence demonstrating that the effect is voltage-induced is provided in Fig. S2.

The thermal stability factor is defined as:

$$\Delta_{AP,P} = \ln(\frac{\tau_{AP,P}}{\tau_0}) = \frac{E_b}{k_B T}\left(1 \pm \frac{H_z - H_s}{H_K^{eff}}\right)^2, \tag{4}$$

from which we obtain the average stability factor:

$$\Delta_{AVE} = \frac{\Delta_{AP} + \Delta_P}{2} = \frac{E_b}{k_B T}\left(1 + \left(\frac{H_z - H_s}{H_K^{eff}}\right)^2\right), \tag{5}$$

Fig. 3(c) presents $\Delta_{AVE}$ as a function of $H_z$ under various currents. The curves exhibit a parabolic shape with positive curvature, consistent with the Eq. (5). The magnetic field at each minimum point, where $H_z = H_s$, aligns with the AP rate = 50% points in Fig. 3(a), demonstrating the VCEC effect. The shift of $\Delta_{AVE}$ indicates that the MTJ becomes less stable at a positive current, consistent with the VCMA picture. According to VCMA picture, as the current in the positive direction increased, electrons are depleted from the free layer towards the barrier, resulting in decreased anisotropy. We use the $\Delta_{AVE}$ at $H_z = H_s$ ($\Delta_{AVE,min}$) as an indicator of the intrinsic average stability under each current. Fig. 3(d) displays $\Delta_{AVE,min}$ and the corresponding switching rate as a function of the applied current. Unlike the STT scenario, where a current-even contribution from the thermal effect is expected, we observe a linear relationship between $\Delta_{AVE,min}$ and current, confirming the VCMA and suggesting minimal thermal effects. By fitting the slope of this trend, the VCMA is estimated to be 2.43 fJ/Vm, consistent with the reported value of the Ta buffer layer[46].

Generally, the exponential parameter $n_H$ in Eq. (1) may vary from 1.5 to 2 under different currents. However, this variation does not affect the extracted VCEC and VCMA value. We set $n_H = 2$ to simplify the expression in the above derivative. With the obtained VCEC and VCMA value, both the output level and the switching rate can be predicted based on the bias across the pillar.

The SOT from buffer layer beneath the MTJ pillar has been identified as another method to manipulate stochastic behavior. In our stack, the buffer layer consists of two heavy metal layers with spin Hall angles with opposite signs, reported to be capable of generating a nonzero SOT effective magnetic field in the out-of-plane direction[47]. To examine how the SOT current influences voltage controlled MTJ's stochastic behavior, we performed measurement while applying a current through the spin Hall channel. The phase diagrams of AP rate with an SOT current of 0, -0.3 and -0.5 mA are plotted in Fig. 4(a-c). Interestingly, despite the presence of an extra layer between the free layer and the SOT channel, the SOT-induced effective magnetic field still affects the free layer. The boundary between the AP and P regions, where the AP rate equals 0.5, shifts with the application of the SOT current. Aside from this shift, both the magnetic field and VCEC maintain independent control, remaining unaffected by the additional SOT current.

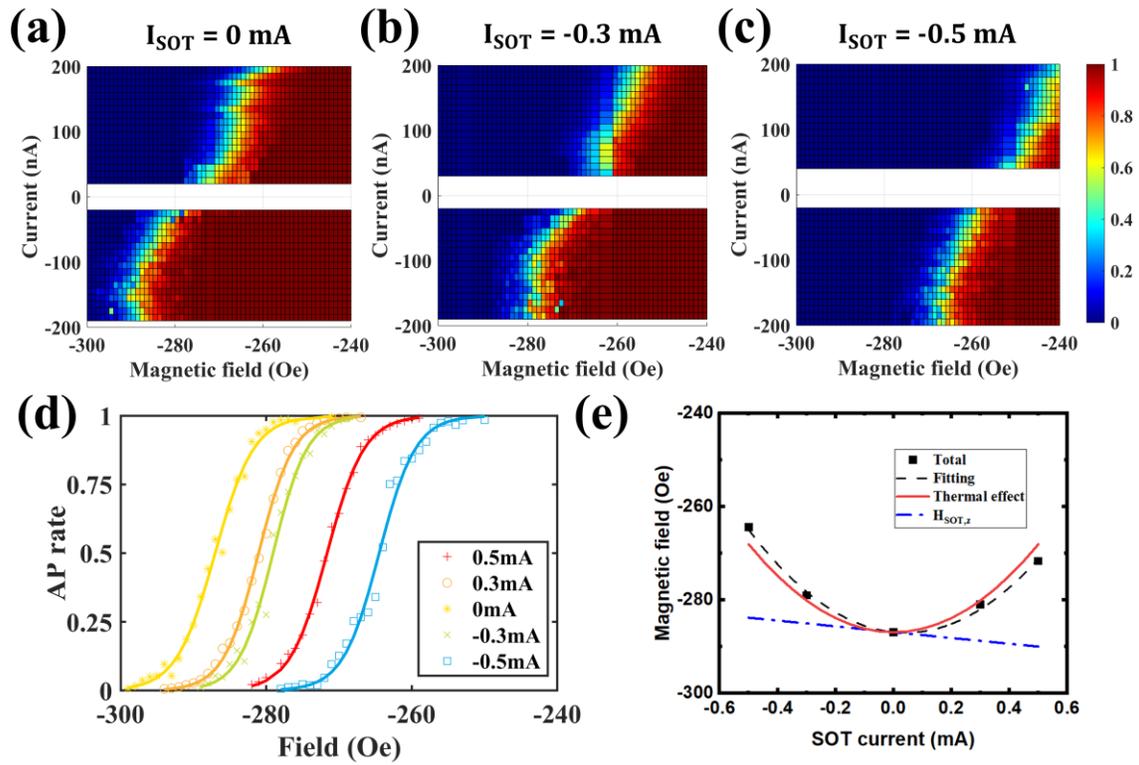

Figure 4. SOT current effect on the AP rate. (a) Phase diagram of the AP rate with an SOT current of 0 mA, (b) with an SOT current of -0.3 mA, and (c) with an SOT current of -0.5 mA, showing the AP rate under varying magnetic fields and applied currents. Each data point is derived from a 40-second waveform. (d) Influence of SOT current induced effect on the AP rate at I = -100 nA. (e) Corresponding magnetic field at AP = 0.5, plotted as a function of SOT current, with the SOT effect separated into spin-orbit torque (SOT) and thermal contributions.

To study the SOT effective field, the relationship between the AP rate and the magnetic field under different SOT current, with a fixed current through the pillar (-100 nA), is shown in Fig. 4(d). The magnetic field at AP rate = 0.5 for various SOT currents is plotted in Fig. 4(e). The primary effect of the SOT current on the effective field is proportional to the square of the current, indicating a Joule heating effect. A minor linear component suggests a non-zero z direction effective magnetic field contribution from the SOT effect, with an efficiency of -6.20 Oe/mA. These two effects are successfully separated through fitting, as shown in Fig. 4(e). The change of the external field by temperature is attributed to the nature of the exchange coupling[48]. To confirm this, we lowered the temperature and measured the R–H loops from 75 K to 275 K under different voltages. The temperature dependence of the stray magnetic field was extracted through fitting and is plotted in Fig. S3(d). We observed that as the temperature increases, the slope of the stray field also increases, reaching 2.26 Oe/K near room temperature. For the SOT-induced thermal effect, the change in exchange coupling field was measured to be approximately 20 Oe, corresponding to a temperature increase of 8.9 K at a SOT current of 0.5 mA. The increase in temperature also results in a decrease in the average thermal stability, as shown in Fig. S4(b).

Since the voltage-induced effects consume negligible power, they minimally interfere with SOT-induced effects via thermal influence. Additionally, VCEC shows negligible variation with temperature changes. This property prevents any cross-interference between VCEC and SOT through thermal effects, enabling independent control by voltage and SOT. In practice, the VCEC + SOT scheme is superior to the STT + SOT scheme, not only due to lower power consumption but also because of the independent control it offers.

In conclusion, we fabricated nano-size perpendicular sMTJs and successfully confirmed the existence of VCEC through the observed switching direction. The VCEC enables precise control over the MTJ output level, allowing energy-efficient operation with power consumption as low as 40 nW. Since this is a voltage-induced effect, power consumption could be further reduced by enhancing VCEC efficiency with no apparent fundamental limitations. Even lower power is anticipated with a thicker and higher-quality MgO barrier. This low-power mechanism opens possibilities for applying sMTJs in large networks with minimal power requirements. Additionally, manipulation by SOT current has been shown to be feasible without affecting the VCEC, even with the presence of the thermal effect. Although in our case, the SOT current primarily influences exchange bias coupling through thermal effects, a SOT channel with a higher spin Hall angle could provide an additional energy-efficient approach for MTJ control, complementing the VCEC mechanism.

METHODS

The stack structure of Ta (5.0 nm)/Pd (2.0 nm)/[Co (0.3 nm)/Pd (0.7 nm)]$_3$/Co (0.3 nm)/Ru (0.6 nm)/Ta (0.3 nm)/CoFeB (1.0 nm)/MgO (2.0 nm)/CoFeB (1.3 nm)/Ta (0.7 nm)/[Pd (0.7 nm)/Co (0.3 nm)]$_8$ is deposited on a thermally oxidized Si substrate at room temperature in the Shamrock system using DC and RF magnetic sputtering at the base pressure of $5 \times 10^{-8}$ Torr [Fig. 1(a)]. After a rapid thermal annealing (RTA) process at 250°C for 10 minutes, all the magnetic layers exhibit a perpendicular easy axis. The stack is then processed into MTJ pillars using electron-beam lithography and Ar-ion milling. The electrode waveguide is deposited using CHA with Ti (10 nm)/Au (100 nm). Devices of various sizes, with diameters of 100, 150, and 200 nm, are fabricated on the same stack. The 150 and 200 nm devices exhibit higher thermal stability, resulting in fewer switching events within the measurement time window. Consequently, the 100 nm devices are chosen for the following time-dependent stochastic measurement.

Electrical measurements are performed with a constant current provided by current source Keithley 2400. The time-dependent voltage is measured by digital oscilloscope (Tektronix DPO 72004C) after removing the DC component with a bias tee and amplifying the signal with Stanford Research 560. Considering the parasitic capacitance C ~ 5 pF and R ~ 5 MΩ, the RC time constant is estimated to be around 25 ms. All the low temperature R-H measurement is performed in PPMS chamber (Quantum Design, Inc.).

## ASSOCIATED CONTENT

**Supporting Information**.

The following files are available free of charge.

V – I curve of the MTJ pillar and power consumption analyze; correlation analyzes between bias induced field with applied current and voltage; low-temperature minor loop measurements of VCEC and VCMA; SOT effect on MTJ thermal stability factor. (PDF)

## AUTHOR INFORMATION

**Corresponding Author**

Jian-Ping Wang − Department of Electrical and Computer Engineering, University of Minnesota, Minneapolis, Minnesota 55455, United States; Email: jpwang@umn.edu

**Authors**

Qi Jia − Department of Electrical and Computer Engineering, University of Minnesota, Minneapolis, Minnesota 55455, United States

Onri J. Benally − Department of Electrical and Computer Engineering, University of Minnesota, Minneapolis, Minnesota 55455, United States

Brandon Zink − Department of Electrical and Computer Engineering, University of Minnesota, Minneapolis, Minnesota 55455, United States; Present Address: National Institute of Standards and Technology, Gaithersburg, MD, United States

Delin Zhang − Department of Electrical and Computer Engineering, University of Minnesota, Minneapolis, Minnesota 55455, United States; Present Address: School of Electronic and Information Engineering, Tiangong University, China

Yang Lv − Department of Electrical and Computer Engineering, University of Minnesota, Minneapolis, Minnesota 55455, United States

Shuang Liang − Department of Chemical Engineering and Materials Science, University of Minnesota, Minneapolis, MN 55455, USA

Deyuan Lyu − Department of Electrical and Computer Engineering, University of Minnesota, Minneapolis, Minnesota 55455, United States

Yu-Chia Chen − Department of Electrical and Computer Engineering, University of Minnesota, Minneapolis, Minnesota 55455, United States

Yifei Yang − Department of Electrical and Computer Engineering, University of Minnesota, Minneapolis, Minnesota 55455, United States

Yu Han Huang − Department of Materials Science and Engineering, National Yang Ming Chiao Tung University, Hsinchu, Taiwan

**Notes**

The authors declare no competing financial interest.

ACKNOWLEDGMENT

This material is based upon work supported in part by the National Science Foundation 2230963, ASCENT: TUNA: TUnable Randomness for Natural Computing, and seed grant of NSF Minnesota MRSEC center. This work was supported in part by ASCENT, one of six centers in JUMP, a Semiconductor Research Corporation (SRC) program sponsored by DARPA. Portions of this work were conducted in the Minnesota Nano Center, which is supported by the National Science Foundation through the National Nanotechnology Coordinated Infrastructure (NNCI) under Award Number ECCS-2025124.

REFERENCES

(1) Zink, B. R.; Lv, Y.; Wang, J.-P. Review of Magnetic Tunnel Junctions for Stochastic Computing. *IEEE Journal on Exploratory Solid-State Computational Devices and Circuits* **2022**, *8* (2), 173–184.
(2) Fukami, S.; Ohno, H. Perspective: Spintronic Synapse for Artificial Neural Network. *Journal of Applied Physics* **2018**, *124* (15), 151904.
(3) Sengupta, A.; Roy, K. Encoding Neural and Synaptic Functionalities in Electron Spin: A Pathway to Efficient Neuromorphic Computing. *Applied Physics Reviews* **2017**, *4* (4), 041105.
(4) Brown, B. D.; Card, H. C. Stochastic Neural Computation. I. Computational Elements. *IEEE Transactions on Computers* **2001**, *50* (9), 891–905.
(5) Canals, V.; Morro, A.; Oliver, A.; Alomar, M. L.; Roselló, J. L. A New Stochastic Computing Methodology for Efficient Neural Network Implementation. *IEEE Transactions on Neural Networks and Learning Systems* **2016**, *27* (3), 551–564.
(6) Bodiwala, S.; Nanavati, N. An Efficient Stochastic Computing Based Deep Neural Network Accelerator with Optimized Activation Functions. *Int. j. inf. tecnol.* **2021**, *13* (3), 1179–1192.

(7) Daniel, J.; Sun, Z.; Zhang, X.; Tan, Y.; Dilley, N.; Chen, Z.; Appenzeller, J. Experimental Demonstration of an On-Chip p-Bit Core Based on Stochastic Magnetic Tunnel Junctions and 2D MoS2 Transistors. *Nat Commun* **2024**, *15* (1), 4098.

(8) Debashis, P.; Faria, R.; Camsari, K. Y.; Appenzeller, J.; Datta, S.; Chen, Z. Experimental Demonstration of Nanomagnet Networks as Hardware for Ising Computing. In *2016 IEEE International Electron Devices Meeting (IEDM)*; 2016; p 34.3.1-34.3.4.

(9) Zand, R.; Camsari, K. Y.; Datta, S.; Demara, R. F. Composable Probabilistic Inference Networks Using MRAM-Based Stochastic Neurons. *J. Emerg. Technol. Comput. Syst.* **2019**, *15* (2), 17:1-17:22.

(10) Borders, W. A.; Pervaiz, A. Z.; Fukami, S.; Camsari, K. Y.; Ohno, H.; Datta, S. Integer Factorization Using Stochastic Magnetic Tunnel Junctions. *Nature* **2019**, *573* (7774), 390–393.

(11) Daniels, M. W.; Madhavan, A.; Talatchian, P.; Mizrahi, A.; Stiles, M. D. Energy-Efficient Stochastic Computing with Superparamagnetic Tunnel Junctions. *Phys. Rev. Appl.* **2020**, *13* (3), 034016.

(12) Mizrahi, A.; Hirtzlin, T.; Fukushima, A.; Kubota, H.; Yuasa, S.; Grollier, J.; Querlioz, D. Neural-like Computing with Populations of Superparamagnetic Basis Functions. *Nat Commun* **2018**, *9* (1), 1533.

(13) Liyanagedera, C. M.; Sengupta, A.; Jaiswal, A.; Roy, K. Stochastic Spiking Neural Networks Enabled by Magnetic Tunnel Junctions: From Nontelegraphic to Telegraphic Switching Regimes. *Phys. Rev. Applied* **2017**, *8* (6), 064017.

(14) Sengupta, A.; Parsa, M.; Han, B.; Roy, K. Probabilistic Deep Spiking Neural Systems Enabled by Magnetic Tunnel Junction. *IEEE Transactions on Electron Devices* **2016**, *63* (7), 2963–2970.

(15) Srinivasan, G.; Sengupta, A.; Roy, K. Magnetic Tunnel Junction Enabled All-Spin Stochastic Spiking Neural Network. In *Design, Automation & Test in Europe Conference & Exhibition (DATE), 2017*; IEEE: Lausanne, Switzerland, 2017; pp 530–535.

(16) Faria, R.; Camsari, K. Y.; Datta, S. Implementing Bayesian Networks with Embedded Stochastic MRAM. *AIP Advances* **2018**, *8* (4), 045101.

(17) Camsari, K. Y.; Faria, R.; Sutton, B. M.; Datta, S. Stochastic p -Bits for Invertible Logic. *Phys. Rev. X* **2017**, *7* (3), 031014.

(18) Misra, S.; Bland, L. C.; Cardwell, S. G.; Incorvia, J. A. C.; James, C. D.; Kent, A. D.; Schuman, C. D.; Smith, J. D.; Aimone, J. B. Probabilistic Neural Computing with Stochastic Devices. *Advanced Materials* **2023**, *35* (37), 2204569.

(19) Camsari, K. Y.; Debashis, P.; Ostwal, V.; Pervaiz, A. Z.; Shen, T.; Chen, Z.; Datta, S.; Appenzeller, J. From Charge to Spin and Spin to Charge: Stochastic Magnets for Probabilistic Switching. *Proceedings of the IEEE* **2020**, *108* (8), 1322–1337.

(20) Safranski, C.; Kaiser, J.; Trouilloud, P.; Hashemi, P.; Hu, G.; Sun, J. Z. Demonstration of Nanosecond Operation in Stochastic Magnetic Tunnel Junctions. *Nano Lett.* **2021**, *21* (5), 2040–2045.

(21) Hayakawa, K.; Kanai, S.; Funatsu, T.; Igarashi, J.; Jinnai, B.; Borders, W. A.; Ohno, H.; Fukami, S. Nanosecond Random Telegraph Noise in In-Plane Magnetic Tunnel Junctions. *Phys. Rev. Lett.* **2021**, *126* (11), 117202.

(22) Schnitzspan, L.; Kläui, M.; Jakob, G. Nanosecond True-Random-Number Generation with Superparamagnetic Tunnel Junctions: Identification of Joule Heating and Spin-Transfer-Torque Effects. *Phys. Rev. Applied* **2023**, *20* (2), 024002.

(23) Kaiser, J.; Rustagi, A.; Camsari, K. Y.; Sun, J. Z.; Datta, S.; Upadhyaya, P. Subnanosecond Fluctuations in Low-Barrier Nanomagnets. *Phys. Rev. Appl.* **2019**, *12* (5), 054056.

(24) Soumah, L.; Desplat, L.; Phan, N.-T.; Valli, A. S. E.; Madhavan, A.; Disdier, F.; Auffret, S.; Sousa, R.; Ebels, U.; Talatchian, P. Nanosecond Stochastic Operation in Perpendicular Superparamagnetic Tunnel Junctions. arXiv February 5, 2024.

(25) Capriata, C. C. M.; Chaves-O'Flynn, G. D.; Kent, A. D.; Malm, B. G. Enhanced Stochastic Bit Rate for Perpendicular Magnetic Tunneling Junctions in a Transverse Field. In *2023 International Conference on Noise and Fluctuations (ICNF)*; 2023; pp 1–4.

(26) Rehm, L.; Capriata, C. C. M.; Misra, S.; Smith, J. D.; Pinarbasi, M.; Malm, B. G.; Kent, A. D. Stochastic Magnetic Actuated Random Transducer Devices Based on Perpendicular Magnetic Tunnel Junctions. *Phys. Rev. Applied* **2023**, *19* (2), 024035.

(27) Zink, B. R.; Wang, J.-P. Influence of Intrinsic Thermal Stability on Switching Rate and Tunability of Dual-Biased Magnetic Tunnel Junctions for Probabilistic Bits. *IEEE Magnetics Letters* **2021**, *12*, 1–5.

(28) Zink, B. R.; Lv, Y.; Wang, J.-P. Independent Control of Antiparallel- and Parallel-State Thermal Stability Factors in Magnetic Tunnel Junctions for Telegraphic Signals With Two Degrees of Tunability. *IEEE Transactions on Electron Devices* **2019**, *66* (12), 5353–5359.

(29) Zink, B. R.; Lv, Y.; Wang, J.-P. Telegraphic Switching Signals by Magnet Tunnel Junctions for Neural Spiking Signals with High Information Capacity. *Journal of Applied Physics* **2018**, *124* (15), 152121.

(30) Lv, Y.; Bloom, R. P.; Wang, J.-P. Experimental Demonstration of Probabilistic Spin Logic by Magnetic Tunnel Junctions. *IEEE Magn. Lett.* **2019**, *10*, 1–5.

(31) Zand, R.; Camsari, K. Y.; Pyle, S. D.; Ahmed, I.; Kim, C. H.; DeMara, R. F. Low-Energy Deep Belief Networks Using Intrinsic Sigmoidal Spintronic-Based Probabilistic Neurons. In *Proceedings of the 2018 Great Lakes Symposium on VLSI*; ACM: Chicago IL USA, 2018; pp 15–20.

(32) Shim, Y.; Chen, S.; Sengupta, A.; Roy, K. Stochastic Spin-Orbit Torque Devices as Elements for Bayesian Inference. *Sci Rep* **2017**, *7* (1), 14101.

(33) Ostwal, V.; Appenzeller, J. Spin–Orbit Torque-Controlled Magnetic Tunnel Junction With Low Thermal Stability for Tunable Random Number Generation. *IEEE Magnetics Letters* **2019**, *10*, 1–5.

(34) Debashis, P.; Faria, R.; Camsari, K. Y.; Chen, Z. Design of Stochastic Nanomagnets for Probabilistic Spin Logic. *IEEE Magnetics Letters* **2018**, *9*, 1–5.

(35) Debashis, P.; Chen, Z. Tunable Random Number Generation Using Single Superparamagnet with Perpendicular Magnetic Anisotropy. In *2018 76th Device Research Conference (DRC)*; 2018; pp 1–2.

(36) Chen, Y.-B.; Yang, X.-K.; Yan, T.; Wei, B.; Cui, H.-Q.; Li, C.; Liu, J.-H.; Song, M.-X.; Cai, L. Voltage-Driven Adaptive Spintronic Neuron for Energy-Efficient Neuromorphic Computing. *Chinese Phys. Lett.* **2020**, *37* (7), 078501.

(37) Liu, S.; Kwon, J.; Bessler, P. W.; Cardwell, S. G.; Schuman, C.; Smith, J. D.; Aimone, J. B.; Misra, S.; Incorvia, J. A. C. Random Bitstream Generation Using Voltage-Controlled Magnetic Anisotropy and Spin Orbit Torque Magnetic Tunnel Junctions. *IEEE Journal on Exploratory Solid-State Computational Devices and Circuits* **2022**, *8* (2), 194–202.

(38) Mishra, P. K.; Sravani, M.; Bose, A.; Bhuktare, S. Voltage-Controlled Magnetic Anisotropy-Based Spintronic Devices for Magnetic Memory Applications: Challenges and Perspectives. *Journal of Applied Physics* **2024**, *135* (22), 220701.

(39) Shao, Y.; Duffee, C.; Raimondo, E.; Davila, N.; Lopez-Dominguez, V.; Katine, J. A.; Finocchio, G.; Amiri, P. K. Probabilistic Computing with Voltage-Controlled Dynamics in Magnetic Tunnel Junctions. *Nanotechnology* **2023**, *34* (49), 495203.

(40) Raimondo, E.; Grimaldi, A.; Giordano, A.; Chiappini, M.; Carpentieri, M.; Finocchio, G. Random Number Generation Driven by Voltage-Controlled Magnetic Anisotropy and Their Use in Probabilistic Computing. In *2024 IEEE 24th International Conference on Nanotechnology (NANO)*; 2024; pp 326–330.

(41) Chen, Y.-C.; Jia, Q.; Yang, Y.; Huang, Y.-H.; Lyu, D.; Peterson, T. J.; Wang, J.-P. Enhanced Voltage-Controlled Magnetic Anisotropy and Field-Free Magnetization Switching Achieved with High Work Function and Opposite Spin Hall Angles in W/Pt/W SOT Tri-Layers. *Advanced Functional Materials n/a* (n/a), 2416570.

(42) Zhang, D.; Bapna, M.; Jiang, W.; Sousa, D.; Liao, Y.-C.; Zhao, Z.; Lv, Y.; Sahu, P.; Lyu, D.; Naeemi, A.; Low, T.; Majetich, S. A.; Wang, J.-P. Bipolar Electric-Field Switching of Perpendicular Magnetic Tunnel Junctions through Voltage-Controlled Exchange Coupling. *Nano Lett.* **2022**, *22* (2), 622–629.

(43) Zink, B. R.; Zhang, D.; Li, H.; Benally, O. J.; Lv, Y.; Lyu, D.; Wang, J.-P. Ultralow Current Switching of Synthetic-Antiferromagnetic Magnetic Tunnel Junctions Via Electric-Field Assisted by Spin–Orbit Torque. *Advanced Electronic Materials* **2022**, *8* (10), 2200382.

(44) Lyu, D.; Zhang, D.; Gopman, D. B.; Lv, Y.; Benally, O. J.; Wang, J.-P. Ferromagnetic Resonance and Magnetization Switching Characteristics of Perpendicular Magnetic Tunnel Junctions with Synthetic Antiferromagnetic Free Layers. *Appl. Phys. Lett.* **2022**, *120* (1), 012404.

(45) Funatsu, T.; Kanai, S.; Ieda, J.; Fukami, S.; Ohno, H. Local Bifurcation with Spin-Transfer Torque in Superparamagnetic Tunnel Junctions. *Nat Commun* **2022**, *13* (1), 4079.

(46) Peterson, T. J.; Hurben, A.; Jiang, W.; Zhang, D.; Zink, B.; Chen, Y.-C.; Fan, Y.; Low, T.; Wang, J.-P. Enhancement of Voltage Controlled Magnetic Anisotropy (VCMA) through Electron Depletion. *Journal of Applied Physics* **2022**, *131* (15), 153904.

(47) Ma, Q.; Li, Y.; Gopman, D. B.; Kabanov, Yu. P.; Shull, R. D.; Chien, C. L. Switching a Perpendicular Ferromagnetic Layer by Competing Spin Currents. *Phys. Rev. Lett.* **2018**, *120* (11), 117703.

(48) Li, Y.; Jin, X.; Pan, P.; Tan, F. N.; Lew, W. S.; Ma, F. Temperature-Dependent Interlayer Exchange Coupling Strength in Synthetic Antiferromagnetic [Pt/Co]2/Ru/[Co/Pt]4 Multilayers*. *Chinese Phys. B* **2018**, *27* (12), 127502.